\documentclass[aps,prb,twocolumn,groupedaddress,floatfix,showpacs]{revtex4}
\usepackage{graphicx}
\begin{document}
\title{Flux creep in type-II superconductors: self-organized criticality
approach}
\author{R.G. Mints}
\email[]{mints@post.tau.ac.il}
\homepage[]{http://star.tau.ac.il/~mints}
\author{Ilya Papiashvili}
\affiliation{School of Physics and Astronomy, Raymond and Beverly
Sackler Faculty of Exact Sciences, Tel Aviv University, Tel Aviv
69978, Israel}
\date{\today}
\begin{abstract}
We consider the current density distribution function of a flux
creep regime in type-II superconductors by mapping the flux creep
process to the dynamics of a model with a self-organized
criticality. We use an extremal Robin Hood type model which
evolves to Been's type critical state to treat magnetic flux
penetration into a superconductor and derive an analog of the
current-voltage characteristics in the flux creep region.
\end{abstract}
\maketitle
\section{Introduction}
In the presence of currents vortex structure in a type-II
superconductor is subjected to Lorentz force. The value of this force
per unit length of a vortex is ${\bf f}_L={\bf j}\times\vec{\phi}_0$,
where ${\bf j}$ is the current density, ${\bf B}$ is the magnetic
field, $\vec{\phi}_0=\phi_0\,{\bf B}/B$, and $\phi_0$ is the flux
quantum. The Lorentz force acting on a unit volume of a vortex matter
is therefore given by ${\bf F}_L =n\,{\bf f}_L={\bf j}\times {\bf B}$,
where $n({\bf r})=B({\bf r})/\phi_0$ is the density of the
vortices\cite{de_Gennes, Campbell_Ivetts}.
\par
\begin{figure}[htb]
\includegraphics[width=0.7\hsize]{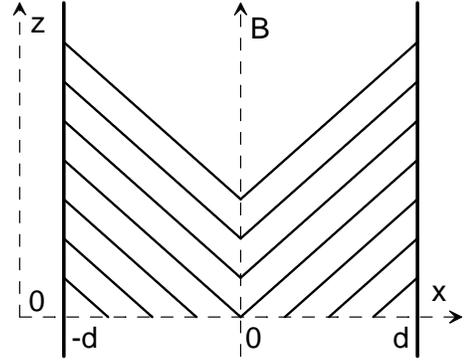}
\caption{Series of Been's critical states in a slab parallel to
the $yz$-plane. A zero-field cooled sample was subjected to a
monotonically increased field parallel to the $z$-axis. The slope
of $B(x)$ is proportional to the critical current density $j_c$.}
\label{fig_01}
\end{figure}

Consider as an illustration a superconducting slab parallel to the
$y,z$-plane and assume that a certain magnetic field $B_a$ is applied
along the $z$-axis as shown in Fig. \ref{fig_01}. In this case we have
${\bf j}=j\,\hat{\bf y}$ and ${\bf B}=B\,\hat{\bf z}$ related by
Maxwell's equation $dB/dx=-\mu_0j$. The Lorentz force $F_L=jB$ is
therefore proportional to the density of vortices $n$ and its gradient
$dn/dx$ as follows from $F_L =j\,B\propto B\,dB/dx\propto n\,dn/dx$.
\par
In high-current density superconductors defects of the crystalline
structure ``pin" the vortices. This pinning leads to creation of
various vortex configurations with flux ``hills'' and flux
``valleys'' between these hills. Indeed, single vortices or
bundles of vortices redistribute spatially if the Lorentz force
$F_L\propto j\propto dn/dx$ overcomes the pinning force $F_{\rm
pin}$. This means that $n\,dn/dx\propto F_{\rm pin}$, {\it i.e.},
a steady vortex structure consists of a composition of fragments
with a certain slope $dn/dx\ne 0$ (the flux hills). On a
macroscopic level we have $F_L\propto j$ and therefore the local
depinning of vortices happens when the local current density $j$
exceeds a certain {\it critical} value $j_c\propto F_{\rm pin}$.
It was pointed out by de Gennes that the flux hills with slopes
statistically fixed by the critical current density $j_c$ are very
much alike to the sand piles\cite{de_Gennes}.
\par
\par
The concept of approaching to flux statics and dynamics in
superconductors with high density of pinning centers was first
suggested by Been\cite{Been_1}. The famous Bean's model assumes that
the current density $j$ is equal to the critical current density $j_c$
everywhere where there are currents. Initially Been considered a
magnetic field independent $j_c$. As a result the spatial increase or
decrease of the field inside the sample is linear. In particular, in a
slab with thickness $2d$ (see Fig.~\ref{fig_01}) the assumption
$j_c={\rm Const}$ leads to $B=B_a-\mu_0j_c|x\mp d|$, where $B_a$ is the
magnetic field at the sample surface. In general, the field dependence
of $j_c$ has to be taken into account in order to give a better
description of the whole wealth of the experimental data available
\cite{Anderson, Kim_Anderson_1}.
\par
In this paper we treat the low-temperature flux creep in type-II
superconductors in the framework of the self-organized criticality
approach\cite{Bak, Jensen, Altshuler, Zaitsev, Wang, Pan, Field,
Aegerter_1, Behnia, Aegerter_2, Mulet}. The distribution function of
the current density $j$ is considered in detail for an extremal Robin
Hood type process \cite{Zaitsev}. We show that this process results in
a self-organized Been's type critical state with a complex dynamics,
which can be mapped to the low-temperature flux creep. This mapping is
used to obtain an analog of a current-voltage characteristics of a
type-II superconductors in the flux creep regime.
\par
The paper is organized as follows. In Sec.~II we discuss the
dynamics of the low-temperature flux creep regime in type-II
superconductors. The self-organized criticality of an extremal
Robin Hood type process is treated in Sec.~III. In Sec.~IV we
introduce the low-temperature flux creep model and derive the
distribution function of the current density $j$. The magnetic
flux penetration into a superconductor and an analog of
current-voltage characteristics of a type-II superconductor in the
flux creep regime are treated in  Sec.~V. Sec.~VI summarizes the
obtained results.
\par
\section{Flux creep in superconductors}
There are few mechanisms resulting in depinning of vortices for
currents with a density less than the critical value. In particular,
both thermally activated depinning and quantum tunnelling
\cite{Anderson, Glazman} result in vortices and bundles of vortices
jumping from one group of pinning centers to another. This type of
vortex motion is called flux creep \cite{Beasley}. The probability of
depinning of a bundle of vortices strongly depends on the current
density and tends to unity when the current density tends to the
critical current density, {\it i.e.}, $j\rightarrow j_c$. The high
level of correlations between vortices in a vortex matter leads to a
very complicated collective behavior of vortices especially in thin
superconducting films where the stray fields result in nonlocality of
the problem \cite{Aranson}. Activation of one vortex can ignite a local
avalanche-type motion of bundles of correlated vortices, {\it i.e.}, in
the flux creep region vortex matter is a system with avalanche driven
dynamics. Such systems with avalanche driven dynamics are the subject
of the modern theory of self-organized criticality which revealed a
variety of power-law distributions for the spatial size and temporal
duration of the avalanches\cite{Bak,Jensen}.
\par
Motion of vortices is accompanied by the field and current energies
dissipation. The heat release during rearrangements (flux avalanches)
in a vortex matter can heat up the whole sample or a part of it to a
temperature higher than the critical temperature $T_c$. Under certain
conditions an avalanche of even a small group of vortices can trigger a
run-away magneto-thermal instability causing the
superconducting-to-normal transition \cite{Mints_Rakhmanov}.
\par
The original Bean's model means that in the critical state the
dependence of $j$ on $E$ is a highly nonlinear stepwise function
\begin{equation}\label{eq_01}
{\bf j}={\bf j}_c\,\cases{0,\quad {\rm if}\quad E=0,\cr 1,\quad {\rm
if}\quad E\ne 0,}
\end{equation}
where ${\bf j}_c=j_c\,{\bf\hat{e}}$ and ${\bf\hat e}={\bf E}/E$ is a
unit vector parallel to the electric field ${\bf E}$ induced by flux
motion.
\par
It is well established now that in the narrow vicinity of the critical
current ($|j-j_c|\ll j_c$), {\it i.e.}, in the flux creep regime the
dependence of $j$ on $E$ is a very steep function given by the power
law
\begin{equation}\label{eq_02}
{\bf j}={\bf j}_c\,\Bigl({E\over E_0}\Bigr)^{1/n} \!\!\!,
\end{equation}
where $n\gg 1$ is a parameter and the field $E_0$ defines the
critical current density. It is common to define $j_c$ as the
current density at $E_0=10^{-6}$~Vcm$^{-1}$. It is worth
mentioning that for $n\gg 1$ we can rewrite Eq. (\ref{eq_02}) in
the following logarithmic form
\begin{equation}\label{eq_03}
{\bf j}={\bf j}_c +{{\bf j}_c\over n}\,\ln\left({E\over
E_0}\right),
\end{equation}
where the omitted terms are of order $1/n^2\ll 1$.
\par
The dependence of $j$ on $E$ given by Eq. (\ref{eq_02}) was first
derived in the framework of the Anderson-Kim model
\cite{Kim_Anderson_1} considering the thermally activated uncorrelated
hopping of bundles of vortices. The vortex-glass \cite{Fisher} and
collective creep \cite{Larkin, Feigelman} models result in more
sophisticated dependencies of $j$ on $E$. However, these dependencies
coincide with the one given by Eq.~(\ref{eq_02}) if $j-j_c\ll j_c$. The
recently developed self-organized criticality approach to the critical
state \cite{Wang,Pan} also leads to Eq.~(\ref{eq_02}) if $j-j_c\ll
j_c$. In the interval $j-j_c\ll j_c$ the power law (\ref{eq_02}) is in
good agreement with numerous experimental data\cite{Gurevich}.
\par
A detailed study of the dynamics of vortices in the flux creep regime
was performed by Field and coworkers\cite{Field}. In their experiments
the magnetic field outside a tubular superconducting sample is ramped
slowly, driving the flux into the tubes outer wall. After the flux
front reaches the inner wall of a tube it spills out into the tube's
interior. The entrance of flux into the tube's interior was detected in
real time. This experiment distinguished between flux leaving the
superconductor in discrete bundles or avalanches. It was shown that the
probability $D(s)$ of an avalanche containing $s$ vortices is a
power-law dependence extending over $1.6$ decades.
\par
Been's critical state dynamics is typical to some other spatially
extended dynamical systems. High number of degrees of freedom in these
systems introduces the problem of treating the effect of coupling
between the individual degrees of freedom. In many cases even very
complicated systems ``self-organize'' so that their description reduces
to a small number of collective degrees of freedom\cite{Bak,Jensen}.
\par
In some dynamical systems the individual degrees of freedom keep each
other in a stable balance, which cannot be described as a
``perturbation" of some decouple state as well as in terms of a small
amount of collective degrees of freedom. This type of the
self-organized systems has to be quite robust. In the opposite case
these systems would not be able to evolve to a stable balanced
``critical" state\cite{Bak,Jensen}. The sand piles and flux hills in
superconductors demonstrate many features typical for the
self-organized critical state.
\par
Several different models were suggested to illustrate and study
dynamical systems with extended spatial degrees of freedom and many
metastable states. These systems evolve into a {\it self-organized}
critical state without a detailed specification of the initial
conditions, i.e., the critical state is an attractor of their dynamics
which is robust with respect to variations of parameters and the
presence of quenched randomness.
\par
\section{Self-organized criticality of extremal processes}
We consider now a specific subclass of {\it extremal} processes
demonstrating the self-organized criticality. In extremal models at
each step only the sites satisfying a certain extremal criterion are
involved into the system dynamics. In particular, the problem of
low-temperature flux creep in superconductors can be mapped to the
Robin Hood type extremal model\cite{Zaitsev}.
\par
\subsection{Robin Hood processes.}
The following picture illustrates the main idea of the Robin Hood type
process. The famous robber leaves everyday his forest for a nearest
market and tries to obtain justice. He finds the richest seller, takes
from him some random amount of goods or money and distributes it among
neighbors. He takes nothing to himself so that overall amount of goods
on the market is conserved. The market reaches a critical state after
some transition time.
\par
Many physical systems with a dynamical variable matching a certain
conservation relation can be treated in the framework of the Robin Hood
model. In particular, the low-temperature flux creep, motion of
dislocations through a random array of point obstacles, domain walls in
ferromagnets, grain boundaries in polycrystals, {\it etc}, can be
mapped to this type of extremal processes\cite{Zaitsev}. We will use
the Robin Hood type model to treat the self-criticality of the
low-temperature flux creep in superconductors in detail.
\par
\subsection{Low-temperature flux creep}
If the temperature is very low than vortices in sites with even
slightly different current density have strongly different probability
of depinning. Therefore, the depinning of vortices will happen first in
the place with the highest current density.
\par
\begin{figure}[htb]
\includegraphics[width=0.85\hsize]{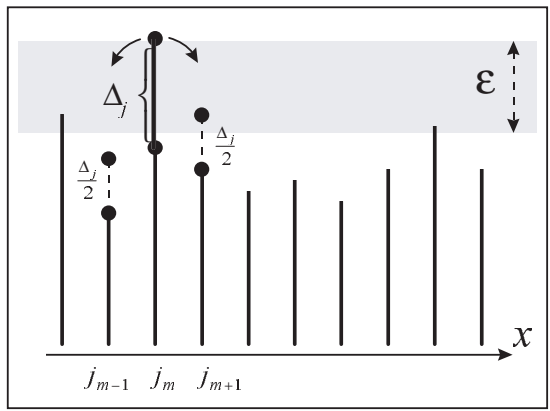}
\caption{Illustration of the dynamics rules of the Robin Hood type
low-temperature flux creep model.}
\label {fig_02}
\end{figure}
\par
The model suggested by Zaitsev \cite{Zaitsev} simulates a part of a
superconductor with $L$ sites using the closed boundary conditions.
Numbers $j_i$ ($0\leq i \le L$) describe the values of the current
density $j$ along the $x$-axis (the current flows along the $y$-axis).
At each simulation step a site number $m$ with the maximal current
density $j_m$ is found (see Fig.~\ref{fig_02}). This current density is
reduced by $\Delta_j$ chosen from a uniform distribution $U(j)$
\begin{equation}\label{eq_04}
j_m \rightarrow j_m-\Delta_j,\qquad j_{m\pm 1} \rightarrow j_{m\pm
1}+\Delta_j/2.
\end{equation}
It is worth mentioning that Zaitsev's model conserves the total current
\begin{equation}\label{eq_05}
I=\sum_{n=1}^L j_n=\langle j\rangle\,L,
\end{equation}
which means that the average value of the critical current density
$\langle j\rangle$ is also conserved. The system approaches the
critical state starting from any initial state with a given value of
$\langle j\rangle$. In this critical state almost all sites have
current densities less than a certain value $j_c$. Each site of the
system maps an area in a superconductor containing a large amount of
vortices but depinning is referred to a single vortex. Motion of this
vortex leads to a rearrangement of a cluster of correlated vortices and
changes the local current density by some value which is determined by
the uniform distribution $U(j)$.
\par
The distribution of the values of $\Delta_j$ which will be referred
below as $\Delta_j$-distribution, defines the number of vortices
participating in an avalanche starting from a depinning of a single
vortex, or, to be more precise, it is related to the number of vortices
leaving a given site as a result of an avalanche. At sufficiently low
temperature this vortex rearrangement is a very fast process comparing
to the thermally activated depinning. Therefore, all the details
concerning these avalanches, including microscopic vortex dynamics, are
``hidden" in $\Delta_j$-distribution.
\par
\subsection{Dynamics of extremal processes}
The fact that the flux creep dynamics conserves the total current $I$
leads to a high level of correlation between the sites. In the extremal
models the extremal sites are located according to some
problem-specific criterion and they provoke changes of the neighbor
sites, again, according to rules specific for a given model. We will
call these sites {\it ignition} sites. The sites drawn in the activity
by the ignition sites will be called {\it involved} sites. An involved
site in its order can became a ignition sites if it matches a certain
extremal criterion. A system in a critical state is characterized by a
critical value of the dynamical parameter. In particular, for the flux
creep model this parameter is $j_c$ which is the least value of $j$ in
the ignition sites. That means that for the flux creep model all
ignition sites have values greater than $j_c$.
\par
\begin{figure}[htb]
\includegraphics[width=0.85\hsize]{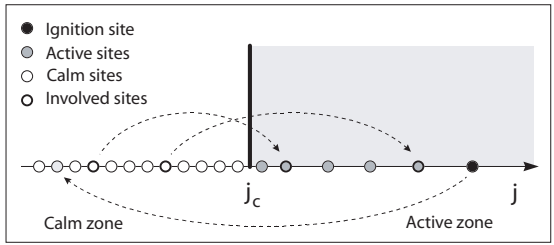}
\caption{The calm and the active zones for a one-dimensional flux
creep model. The rules of dynamics given by Eq. (\ref{eq_04}) are
illustrated by the dashed lines.}
\label {fig_03}
\end{figure}

The rules of dynamics given by Eq. (\ref{eq_04}) are illustrated
schematically in Fig.~\ref{fig_03}. The interval of values of current
densities in the ignition sites we call {\it active} zone. Each site in
the active zone, {\it active} site, at a certain moment becomes an
ignition site. At any step only a small part of all sites is active. As
shown in Fig.~\ref{fig_03} the majority of the sites belongs to the
{\it calm} zone. We will consider the dynamical properties of the
low-temperature creep model (\ref{eq_04}) using this terminology.
\par
\par
\section{Low temperature creep model}
\begin{figure}[htb]
\includegraphics[width=0.85\hsize]{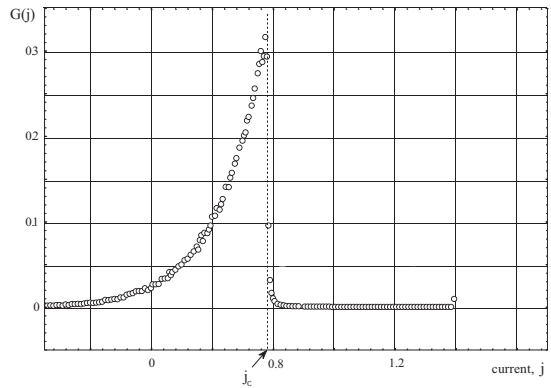}
\caption{The dependence of the distribution function $G(j)$ on the
current density $j$.}
\label{fig_04}
\end{figure}

\begin{figure}[htb]
\includegraphics[width=0.85\hsize]{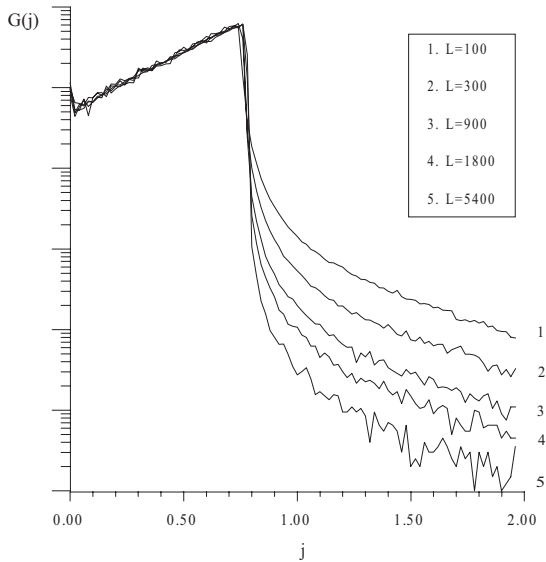}
\caption{Distribution functions $G(j)$ for different values of the
length of the system $L$ in the region $j>j_c$ [``tails'' of $G(j)$].}
\label{fig_05}
\end{figure}

The results of our numerical simulations of a system obeying the
dynamics rules given by Eq. (\ref{eq_04}) demonstrate that $G(j)$
is described by an exponent up to a certain critical value $j_c$,
{\it i.e.},
\begin{equation}\label{eq_06}
G(j)=A\,\exp(j/j_e),\qquad \textrm{for\ \ } j_c-1<j<j_c,
\end{equation}
where $A$ and $j_e$ are parameters of the distribution (see Fig.
\ref{fig_04}). The distribution function $G(j)$ has a sharp cut
off at $j=j_c$. In the active zone $(j>j_c)$ the ``tail'' of
$G(j)$ decreases as $1/L$, when $L\rightarrow\infty$ (see Fig.
\ref{fig_05}).
\par
The distribution function $G(j)$ has another cut off at $j_c-1$ if
the $\Delta_j$-distribution in the interval $(0,1)$ is chosen to
be uniform. Indeed, there are only two options to create a site
with a current density $j$: (a) to decrease the current density in
the ignition site by $\Delta_j$ and (b) to increase the current
density in one of the involved sites by $\Delta_j/2$. The lowest
value of $j$ is obtained by subtracting from the minimum value of
the current density in an ignition site, which is $j_c$, the
maximum value from $U$, which is $1$. As a result the distribution
function $G(j)$ has a cut off from the left. This cut off is not
universal and depends on the $\Delta_j$-distribution. A
continuously decreasing $\Delta_j$-distribution eliminates both
the left cut off and the deviation of $G(j)$ from the exponential
function in the vicinity of $j_c$. The details of a specific
$\Delta_j$-distribution affect only the tail of the function
$G(j)$, {\it i.e.}, its behavior in the interval $j>j_c$.
Therefore, we conclude that the values of the current density $j$
are distributed according to the exponential law given by Eq.
(\ref{eq_06}) in almost all sites of the system.
\par
\par
\subsection{Relation between the parameter $j_e$ and $j_c$}
The relation between the parameter $j_e$ and the critical current
density $j_c$ can be calculated analytically using the
normalization and average current density conservation rules.
Indeed, normalization of the distribution function $G(j)$ given by
Eq. (\ref{eq_06})
\begin{equation}\label{eq_07}
\int_{-\infty}^{j_c}G(j)\,dj=A\,\int_{-\infty}^{j_c} \exp(j/j_e)\,dj=1
\end{equation}
relates $A$, $j_c$, and $j_e$
\begin{equation}\label{eq_08}
A={1\over j_e}\,\exp(-j_c/j_e).
\end{equation}
Next, we calculate the average value of the current density
\begin{equation}\label{eq_09}
\langle j\rangle =\displaystyle{\int_{-\infty}^{j_c}}j\,G(j)\,dj
\end{equation}
and obtain the dependence of $A$, $j_c$, and $j_e$ on $\langle
j\rangle$
\begin{equation}\label{eq_10}
\langle j\rangle=A\,j_e^2\,(j_c/j_e-1)\exp(j_c/j_e).
\end{equation}
\par
Combining Eqs. (\ref{eq_08}) and (\ref{eq_10}) we find the
relation between the critical current density $j_c$, parameter of
the exponent $j_e$, and average current density $\langle j\rangle$
\begin{equation}\label{eq_11}
j_c-j_e=\langle j\rangle.
\end{equation}
\par
If $\Delta_j$-distribution is uniform, then we have $\langle j\rangle
=1/2$ and we find
\begin{equation}\label{eq_12}
j_c-j_e=1/2.
\end{equation}
\par
\par
Our numerical simulations show that the exponential form
(\ref{eq_06}) of the distribution function $G(j)$ is universal,
meaning that it does not depend on the spatial dimension of the
model, on the form of the $\Delta_j$-distribution, on the dynamic
rules of redistribution of $\Delta_j$ between neighbors, and on
the decision on how many sites from the vicinity of an ignition
site are involved in the redistribution process (for example, next
nearest neighbors can be included into the dynamics). Therefore,
we conclude that the exponential behavior of $G(j)$ is typical for
extremal models with short range interactions and a conservation
relation for the dynamical variable.
\par
\subsection{Origin of the exponent}
We performed numerical simulations for different extremal models and
obtained the $G(j)$ distribution for all of them. In addition to the
uniform distribution for $\Delta_j$ we tested also the exponential,
gaussian, and power law distributions. These $\Delta_j$-distributions
are more ``natural" than $U(j)$, since they decay gradually and have no
cutoffs, unlike the uniform distribution. All of them lead to the
exponential form of $G(j)$ if two conditions are satisfied. (a)
Interactions in the model are local, meaning that an activated site
affects only sites in its surrounding of a finite size; (b) At each
step the sum of the dynamical variables ($j_i$) stays constant.
\par
The exponential behavior of $G(j)$ can be obtained analytically as the
most probable distribution of the dynamical variable $j_i$. To simplify
the derivation we use discrete values of $j$ dividing the whole
interval of values of $j$ into small intervals. In this case each site
is characterized by a certain value $j_i$. We denote the number of
sites with the same value $j_i$ as $n_i$ and write the conservation
relations in the form
\begin{equation}\label{eq_13}
\sum_i n_i = L,\qquad\sum_s n_i j_i = \langle j_i\rangle\,L.
\end{equation}
\par
Distribution of $L$ sites between the intervals with $j=j_i$ is
described by a set \{$n_i$\} of numbers of sites in each interval. The
number of states corresponding to the same set \{$n_i$\} is given by
\begin{equation}\label{eq_14}
\Gamma={\frac{L!}{{\prod_i n_i!}}}.
\end{equation}
Using Stirling's formula we have for $\sigma=\ln \Gamma $
\begin{equation}\label{eq_15}
{\sigma=L \ln L - \sum_i n_i \ln n_i}
\end{equation}
The maximum of $\sigma$ can be found using Lagrange method for
conditional extremum with two conditions (\ref{eq_13})
\begin{eqnarray}\label{eq_16}
F&=&\sigma+(\alpha+1)L-\beta\langle j\rangle\,L\nonumber\\
{\frac{\partial{F}}{\partial{n_i}}}&=&-\ln n_i +\alpha + \beta j_i=0
\end{eqnarray}
and we get finally
\begin{equation}\label{eq_17}
n_i=\displaystyle{e^{\alpha+\beta j_i}}
\end{equation}
with $\alpha$ and $\beta$ defined by
\begin{equation}\label{eq_18}
\sum_i e^{\alpha+\beta j_i}=L, \qquad \sum_i j_i e^{\alpha+\beta
j_i}=\langle j\rangle\,L.
\end{equation}
Using the continuous form of Eq. (\ref{eq_18})
\begin{equation}\label{eq_19}
\int  e^{\alpha+\beta j}=L, \qquad \int  j\, e^{\alpha+\beta
j}dj=\langle j\rangle\,L
\end{equation}
we find
\begin{equation}\label{eq_20}
\beta =\frac{1}{j_e}, \qquad e^\alpha=A=\frac{1}{j_e} \exp(-j_c/j_e),
\end{equation}
{\it i.e.}, the values $j_c$, $j_e$, and $\langle j\rangle$ are
related by Eq. (\ref{eq_11}) and
\begin{equation}\label{eq_21}
G(j)={1\over j_e}\exp\left({j-j_c \over j_e}\right)= {1\over
{e\,j_e}}\exp\left({j-\langle j\rangle \over j_e}\right).
\end{equation}
\par
The distribution function given by Eq. (\ref{eq_21}) contains only
one independent parameter, $j_e$ which is characterizing the width
of the distribution function $G(j)$. The value of $j_e$ is
proportional to the width $\delta\Delta_j$ of the
$\Delta_j$-distribution ($j_e\propto\delta\Delta_j$) and depends,
in particular, on the spatial dimension of the model, on the
number of the neighbors of the ignition site, etc.
\par
It is worth mentioning that the main features of the distribution
function $G(j)$ can be formulated in terms of a certain effective
``temperature" associated with the system. Indeed, the above
derivation of the distribution function $G(j)$ is based on the
same arguments which are used to derive the Gibbs distribution for
a system of nonidentical particles. We can write Eq. (\ref{eq_21})
in the form $G(j)=\exp (-\epsilon_j/\tau)/j_e$, where the
effective ``energy" $\epsilon_j$ and ``temperature" $\tau$ are
defined as $\epsilon_j=a\,(j_c-j)$ and $\tau =\alpha j_e$ and $a$
is an arbitrary coefficient. Using Eq. (\ref{eq_11}) we find for
the average energy
\begin{equation}\label{eq_22}
\langle\epsilon_j\rangle=a\,(j_c-\langle j\rangle)=a\,j_e=\tau.
\end{equation}
The value of the parameter $j_e$ is proportional to the width
$\delta\Delta_j$ of the $\Delta_j$-distribution, and therefore the
effective temperature $\tau\propto\delta\Delta_j$.
\par
\section{Flux penetration model}
Extremal models can be useful for studying macroscopic processes in
superconductors and, in particular, the process of penetration of
magnetic flux. In this section we modify our model to allow for
considering this process. The periodic boundary conditions which we
used above simulated an inner part of the system located far from the
sample edges.
\par
Consider now a slab subjected to a magnetic field parallel to its
surface as shown in Fig.~\ref{fig_01}. The symmetry of the problem
allows to perform the numerical simulations in one half of the sample.
Assume that the applied field rises up with a certain rate $\dot{h}$
and the flux penetrates inside the sample from its left ($x=-d$) edge.
In this case the boundary condition at $x=-d$ takes the form
\begin{equation}\label{eq_23}
j_0(t+\delta{t}) = j_0(t) + h\,\delta{t},
\end{equation}
where $j_0(t)$ is the current density at $x=-d$.
\par
The dynamics rule at the middle of the slab ($x=0$) has the form
\begin{eqnarray}\label{eq_24}
j_L(t+\delta t)&=&j_L(t)-\Delta_j,\nonumber\\
j_{L-1}(t+\delta t)&=&j_{L-1}(t)+\Delta_j/2,
\end {eqnarray}
where $j_L(t)$ is the current density at $x=0$. It is worth mentioning
that there is no current conservation at the sample boundaries at each
simulation step. However, in a stationary state the current
conservation holds in average for large time intervals.
\par
\subsection {Introduction of the ``real'' time scale}
The above self-organized criticality approach to the process of
the low-temperature flux creep formulates the dynamical rules of
the current density dynamics and operates with the simulation
steps only. These steps correspond to the sequential depinning
events in the process of numerical simulation of the model. The
``real" time elapsed between two successive steps in the numerical
simulations can vary significantly. This time mismatch has to be
taken into consideration in order to relate the results of the
calculations and experiments.
\par
The temporal variation of the applied magnetic field has its own
time-scale, which has to be synchronized with the ``inner clock"
of the numerical simulations. This synchronization can be done by
calculating the ``real" time interval between two successive
depinning events. The extremal models of the low-temperature creep
are based on the assumption that the depinning probability $P_d$
strongly depends on the maximal current density $j_m$ since the
higher is $j_m$ the higher is the depinning probability. We assume
that this dependence has the ``natural'' exponential form
\begin{equation}\label{eq_25}
P_d\propto\exp\left[{j_m-j_c\over
j_1}\right]\propto\exp\left[j_m\over j_1\right],\quad {\rm
for}\quad j_m<j_c.
\end {equation}
The mean time between two depinning events $\langle\delta
t\rangle$ is inversely proportional to $P_d$ and therefore we
write
\begin {equation}\label{eq_26}
\langle\delta{t}\rangle (t) = \delta{t}_{\rm
tick}\,\exp\left[-{j_m(t)\over j_1}\right],
\end {equation}
where $\delta{t}_{\rm tick}$ is the time interval corresponding to
one {\it tick} of the ``real" time clock. The last equation
provides a synchronization mechanism between the steps in the
numeric simulations and the ``real" time intervals.
\par
Using Eqs. (\ref{eq_23}) and (\ref{eq_26}) we arrive to the
synchronized boundary condition at $x=-d$ in the form
\begin{equation}\label{eq_27}
j_0(t+\delta t)=j_0(t)+h\,\delta{t_{\rm
tick}}\exp\left[-{j_m(t)\over j_1}\right].
\end{equation}
Changing the time scale, so that $h\,\delta{t_{\rm tick}}
\rightarrow h$ we rewrite Eq. (\ref{eq_27}) in the form convenient
for recursive calculations
\begin{equation}\label{eq_28}
j_0\rightarrow j_0+h\,\exp\left[-{j_m\over j_1}\right].
\end{equation}
\par
\begin{figure}[t]
\includegraphics[width=0.85\hsize]{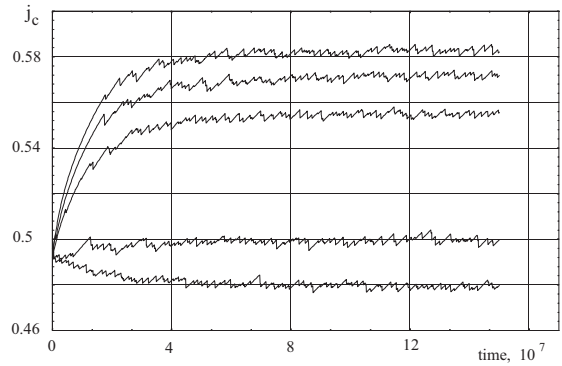}
\caption{Starting from the same initial distribution, the current
density approaches different asymptotic values depending on the
ramp rate $h=4\times 10^{-5}\!,\ 3\times 10^{-5}\!,\ 2\times
10^{-5}\!,\ 5\times 10^{-6}\!,\ 3\times 10^{-6}$ (from top to
bottom).}
\label{fig_06}
\end{figure}
\par
\begin{figure}[htb]
\includegraphics[width=0.85\hsize]{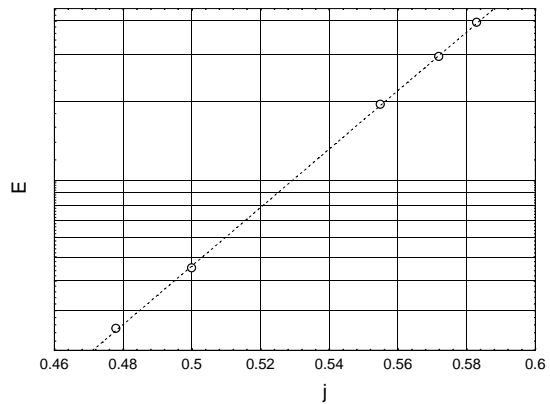}
\caption{Analogue of the current-voltage characteristics.}
\label{fig_07}
\end{figure}
\par
We treat now an analog of the $j$-$E$ curve for the low-temperature
magnetic flux creep. The self-organized criticality model does not
introduce the electrical field $E$ explicitly. Therefore, we have to
extend the above model relating the field $E$ to a certain
characteristics of the model.
\par
When a stable critical state is established, the magnetic flux
redistribute inside the sample keeping the critical current density
$j_c$ almost constant. According to the Faraday law a varying magnetic
field generates an electrical field $E$. The higher is the ramp rate
$h$ of the magnetic field, the higher is an electrical field. We assume
here that the dependence between $E$ and $h$ is linear, {\it i.e.},
$E\propto h$. We demonstrate in Fig. \ref{fig_06} how an asymptotic
current density $j$ is established for several values of the magnetic
field ramp rate $h$. In Fig. \ref{fig_07} is shown the dependence of
$E\propto h$ on the asymptotic value of $j$, which is an analog of a
$j$-$E$ curve of Bean's critical state. This logarithmic $j$-$E$ curve
is consistent with Eq. (\ref{eq_03}) and numerous experimental
data\cite{Gurevich}.
\par
\section{Summary}
We demonstrate that an extremal Robin Hood type model evolves to a
Been's type critical state. The distribution function of the
current density $G(j)$ in this self-organized state was obtained
by numerical simulations as well as analytically. We found that
$G(j)$ has a characteristic cut off at the critical current
density. We map the low-temperature magnetic flux creep process to
dynamics of an extremal model with Been's type critical state to
treat magnetic flux penetration into a superconductor and derive
an analog of the current-voltage characteristics in the flux creep
regime.
\par
\begin{acknowledgments}
This research was supported by The Israel Science Foundation (grant No.
283/00-11.7).
\end{acknowledgments}

\end{document}